%7-12
\def\a{\alpha}\def\b{\beta}\def\d{\delta}\def\e{\epsilon}
\def\f{\phi}\def\g{\gamma}\def\h{\theta}
\def\l{\lambda}\def\m{\mu}\def\n{\nu}\def\o{\omega}\def\q{\psi}\def\r{\rho}
\def\y{\eta}

\def\D{\Delta}\def\L{\Lambda}

\def\de{\partial}
\def\inf{\infty}\def\id{\equiv}\def\ha{{1\over 2}}

\def\ni{\noindent}\def\({\left(}\def\){\right)}\def\[{\left[}\def\]{\right]}
\def\lra{\leftrightarrow}

\def\const{{\rm const}}\def\ex{{\rm e}}
\def\arcth{{\rm arctanh\,}}
\def\arcsh{{\rm arcsinh\,}}

\def\mn{{\mu\nu}}

\def\tran{transformations }\def\coo{coordinates }

\def\ps{phase space }

\def\rep{representation }

\def\pb{Poisson brackets }

\def\ads{anti-de Sitter }
\def\poi{Poincar\'e }
\def\des{de Sitter }
\def\ades{(anti)-de Sitter }\def\QM{quantum mechanics }

\def \schr{Schr\"odinger }

\def\wrt{with respect to }
\def\eom{equations of motion }
\def\cor{commutation relations }
\def\ev{expectation value }\def\bc{boundary conditions }

\def\section#1{\bigskip\noindent{\bf#1}\smallskip}

\def\nota{\footnote{$^\dagger$}}
\font\small = cmr8

\def\PL#1{Phys.\ Lett.\ {\bf#1}}
\def\PRL#1{Phys.\ Rev.\ Lett.\ {\bf#1}}
\def\PR#1{Phys.\ Rev.\ {\bf#1}}\def\CQG#1{Class.\ Quantum Grav.\ {\bf#1}}

\def\JoP#1{J.\ Phys.\ {\bf#1}} \def\IJMP#1{Int.\ J. Mod.\ Phys.\ {\bf #1}}

\def\AoP#1{Ann.\ Phys.\ {\bf#1}}
\def\LMP#1{Lett.\ Math.\ Phys.\ {\bf#1}}
\def\JHEP#1{JHEP\ {\bf#1}}

\def\arx#1{{\tt arXiv:#1}}

\def\ref#1{\medskip\everypar={\hangindent 2\parindent}#1}
\def\beginref{\begingroup
\bigskip
\centerline{\bf References}
\nobreak\noindent}
\def\endref{\par\endgroup}

\input epsf

\def\pres{\sqrt{1-\b^2P_k^2}}
\def\den{\sqrt{1+(\a x_k+\b p_k)^2}}
\def\bra{\langle}\def\ket{\rangle}\def\brah{\langle\hat }
\def\hx{\hat x}\def\hp{\hat p}

\def\cX{{\cal X}}\def\cP{{\cal P}}\def\cE{{\cal E}}
\def\prep{\sqrt{1-\b^2P_r^2}}\def\tP{{\tt\bf P}}
\def\hyp{hypergeometric }
\def\ur{uncertainty relations }
\font\smallmath = cmmi8

\magnification=1200

{\nopagenumbers
\line{}
\vskip20pt
\centerline{\bf Classical and quantum mechanics}
\smallskip
\centerline{\bf of the nonrelativistic Snyder model in curved space}
\vskip20pt
\centerline{{\bf S. Mignemi}\nota{e-mail: smignemi@unica.it}}
\vskip10pt
\centerline {Dipartimento di Matematica, Universit\`a di Cagliari}
\centerline{viale Merello 92, 09123 Cagliari, Italy}
\smallskip
\centerline{and INFN, Sezione di Cagliari}
\vskip60pt
\centerline{\bf Abstract}
\vskip10pt
{\noindent
The Snyder-de Sitter (SdS) model is a generalization of the Snyder model to a spacetime
background of constant curvature. It is an example of noncommutative spacetime admitting
two fundamental scales besides the speed of light, and is invariant under the action of the
de Sitter group.
Here we consider its nonrelativistic counterpart, i.e.\ the Snyder model restricted
to a three-dimensional sphere, and the related model obtained by considering the anti-Snyder
model on a pseudosphere, that we call anti-Snyder-de Sitter (aSdS).

By means of a nonlinear transformation relating the SdS phase space variables to canonical
ones, we are able to investigate the classical and the quantum mechanics of a free particle
and of an oscillator in this framework.
As in their flat space limit, the SdS and aSdS models exhibit rather different properties.
In the SdS case, a lower bound on the localization in position and momentum space arises,
which is not present in the aSdS model. In the aSdS case, instead, a specific combination
of position and momentum \coo cannot exceed a constant value.

We explicitly solve the classical and the quantum equations for the motion of the free
particle and of the harmonic oscillator. In both the SdS and aSdS cases, the frequency of
the harmonic oscillator acquires a dependence on the energy.
}
\vskip80pt
P.A.C.S. Numbers: 02.40.Gh; 45.20.Jj; 03.65.Ca.
\vfil\eject}
\section{1. INTRODUCTION}
Already in 1947, in the attempt to introduce a short distance cutoff in field theory,
Snyder proposed a model of noncommutative spacetime, admitting a fundamental length scale
[1]. In spite of the presence of a fundamental length, the model was invariant under the
action of the Lorentz group.
Snyder's proposal was revived in recent years, when considerations on
quantum gravity and string theory suggested that the structure of spacetime may
be noncommutative at scales close to the Planck length [2].
Noncommutativity of spacetime can in fact account for the modification of
the Heisenberg uncertainty relations [3] and the existence of a minimal bound on the
localization of particles in spacetime [4] that follow from any quantum theory of spacetime.
Actually, it has been argued that already at the Newtonian level the action of gravity calls
for a modification of the \ur [5].

Similar arguments suggest that any theory of quantum gravity should require the introduction
of a fundamental observer-independent scale with the physical dimension of an energy,
in addition to the speed of light.
The new fundamental energy scale sets a bound on the allowed values of the momentum of the
elementary particles, and gives rise to a deformation of the \poi symmetry and of the
energy-momentum dispersion relations.
This line of thought, which is at the basis of Doubly Special Relativity
(DSR) [6], is in some sense dual to that leading to noncommutative geometry,
and in fact the natural spacetime background for DSR is noncommutative.

In particular, the Snyder model can be interpreted as an example of DSR [7], and its
dynamics has been investigated in this context in [8]. Further aspects of the Snyder model
were studied in [9].

Pursuing the DSR idea, it was later proposed that the cosmological constant $\L$
could be introduced as a third fundamental observer-independent parameter in the algebra of
spacetime symmetries [10].
The model based on these assumptions was called Triply Special Relativity (TSR), because
it is based on three fundamental parameters. In the following, we shall usually adopt
the more modest denomination Snyder-\des (SdS) model, especially when dealing with
the nonrelativistic limit.

The proposal of TSR was justified by considerations arising from quantum gravity [10].
From a more formal point of view, its interest relies on the fact that both spacetime
and momentum space are noncommutative, and on its symmetry for the interchange
of positions and momenta, that gives rise to minimal uncertainties for both of them.
In spite of its complexity, the model can be solved exactly in some simple cases,
as we shall show.
\bigskip

TSR is based on the algebra generated by the positions $x_\m$, momenta $p_\m$
and Lorentz generators $J_\mn$. The Lorentz generators satisfy the usual \cor of the
Lorentz algebra, together with
$$[J_\mn,x_\l]=i\hbar(\y_{\m\l}x_\n-\y_{\n\l}x_\m),\qquad
[J_\mn,p_\l]=i\hbar(\y_{\m\l}p_\n-\y_{\n\l}p_\m),\eqno(1.1)$$
while
$$[x_\m,x_\n]=i\hbar\b^2J_\mn,\qquad[p_\m,p_\n]=i\hbar\a^2J_\mn,$$
$$[x_\m,p_\n]=i\hbar\big[\y_\mn+\a^2x_\m x_\n+\b^2p_\m p_\n+\a\b(x_\m p_\n+p_\m x_\n-J_\mn)
\big],
\eqno(1.2)$$
with $\m,\n=0,\dots,3$. The $J_\mn$ and $p_\m$ generate a \des or \ads subalgebra
(depending on the sign of $\a^2$) that describes the spacetime symmetries of the model.
As in DSR, the action of the translations on the spatial \coo is nonlinear.
The coupling constants $\a$ and $\b$ have dimension of inverse length and inverse mass,
respectively.\footnote{$^1$}{We adopt units in which $c=1$. Then $\hbar\sim10^{-38}$g\ cm.}
They are usually identified with the square root of the cosmological constant,
$\a\sim 10^{-24}$cm$^{-1}$,
and with the inverse of the Planck mass, $\b\sim 10^5$g$^{-1}$, but larger scales
compatible with observations may be chosen.
In the following, we require $\a\b\ll1/\hbar$. With the previous identifications, this
inequality is verified by almost 60 orders of magnitude.
The limit $\a\to0$ gives the flat Snyder model, while the limit $\b\to0$ yields
the Heisenberg algebra of quantum mechanics in a de Sitter background endowed with
projective coordinates.

From (1.1) it follows that in TSR both position and momentum components do not commute
among themselves.
For what concerns momenta, this is typical of curved spacetimes, while, as we have seen, the
noncommutativity of positions characterizes Snyder spaces.
Also notable is the duality of the model for position and momentum interchange
through $x_\m\lra p_\m$, $\a\lra\b$. Moreover, as
was recognized in [11], TSR can be viewed as a nonlinear realization of a model introduced
by Yang [12] soon after Snyder's paper, based on the conformal group SO(1,5).
The \cor of Yang's model are again those of the Lorentz algebra, together with
$$[x_\m,x_\n]=i\hbar\b^2J_\mn,\qquad [p_\m,p_\n]=i\hbar\a^2J_\mn,
\qquad[x_\m,p_\n]=i\hbar\y_\mn F,$$
$$[x_\m,F]=-i\hbar\b^2p_\m,\qquad [p_\m,F]=i\hbar\a^2x_\m,\eqno(1.3)$$
where a central charge $F$ that commutes with the Lorentz generators has been introduced.
Although in a quantum setting the linear realization (1.3) can be more convenient, it is not
suitable for an implementation in classical phase space, and we shall not consider it further.

TSR can also be derived from a six-dimensional model, with the higher-dimensional
position and momentum variables satisfying quadratic constraints [13]. Other aspects
of the theory have been studied in [14,15].
\bigskip

Recently, the kinematics and the dynamics of the Snyder model have been investigated
in its three-dimensional (nonrelativistic) version both for classical and quantum systems
[16],\footnote{$^2$}{For a different approach see [17].} exploiting the existence of a
nonlinear map from canonical to Snyder's phase space.
The relevance of the nonrelativistic limit is given by the fact that it exhibits all the
essential features of the original model, but avoids the complications related to
relativistic dynamics.
These investigations evidentiated that the physical properties of the model depend on the
sign of the coupling constant appearing in its algebra.
For positive coupling constant, the momenta are allowed to take any real value, but in the
quantum theory a minimal uncertainty in the positions arises. On the contrary,
in the case of negative coupling constant, that was called anti-Snyder, the momentum is
bounded, but no minimal uncertainty for positions occurs in the quantum theory.

In this paper, we extend the investigations of ref.\ [16] to the case of the
nonrelativistic Snyder model in a constant curvature background.
We carry out this task by exploiting the possibility to
define a simple linear (but of course not canonical) transformation that relates the SdS
algebra to Snyder's.
We  show that, like in the case of the flat Snyder model, the physics depends
strongly on the sign of the coupling constants $\a^2$ and $\b^2$ in (1.2).
In order to obtain a consistent algebra, both constants must have the same sign.
If the sign is positive, the indetermination relations imply the existence of both a
minimal length and a minimal momentum, while for negative sign no restriction arises.

In the first case, formal eigenvalues of position and momentum take discrete values.
However, the existence of a minimal uncertainty implies that the position and momentum
eigenstates cannot have a definite eigenvalue. It follows that the formal solutions of the
eigenvalue equations are not physical, and in fact have diverging values of the
uncertainties $\D x$ and $\D p$.
In a similar situation, Kempf et al.\ [18] suggested that a modified eigenvalue
equation should be introduced that give rise to states that minimize the uncertainty.
In our case we are
not able to find a suitable generalization of their equation, but discuss a class of
functions that give an approximate solution to the problem.

No minimal uncertainties occur instead when $\a^2$ and $\b^2$ are negative. In this
case, however, a specific combination of spatial and momentum \coo must satisfy an
upper bound. This can be interpreted as a sign of a dependence of the geometry on the
energy.

Another interesting application of our method is the study of the harmonic oscillator.
In analogy with the flat Snyder model, both in classical and quantum mechanics its
solution contains corrections of order $(\b^2+\a^2/\o_0^2)E$ to the standard case,
with $E$ the energy and $\o_0$ the classical frequency of the oscillator.
In particular, the frequency of oscillation is no longer independent from the energy.
Furthermore, in the aSdS case the spectrum of energy contains only a finite number of
states.

\section{2. CLASSICAL MECHANICS}
We first consider the implementation of the nonrelativistic SdS model in classical phase
space. The model is defined on a three-dimensional space of constant curvature with
Euclidean signature. Its phase space is endowed with a noncanonical symplectic structure
given by the classical limit of the commutator algebra (1.2). We study the kinematics and
some simple examples of dynamics in one dimension.

\section{2.1 The model}
In classical mechanics, the motion in the nonrelativistic SdS model can be described by
postulating a noncanonical symplectic structure on phase space, with fundamental \pb [13]
$$\eqalign{\{x_i,x_j\}&=\b^2\,J_{ij},\qquad\qquad\{p_i,p_j\}=\a^2\,J_{ij},\cr
\{x_i,p_j\}&=\d_{ij}+\a^2\,x_ix_j+\b^2\,p_ip_j+2\a\b\,p_ix_j,}\eqno(2.1)$$
where $i,j\dots=1,2,3$. The \pb (2.1) are obtained from (1.2) by introducing the standard
expression $J_{ij}=x_ip_j-x_jp_i$ for the generators of the rotations.

In spite of the notation, we allow $\a^2$ and $\b^2$ to be negative,
but, in order for the Jacobi identities to hold, both $\a^2$ and $\b^2$
must have the same sign.\footnote{$^3$}{In this paper, we adopt the following convention:
when $\a^2$ and $\b^2$ are negative, also $\a\b$ is negative, while for expressions linear
in $\a$ and $\b$ we define $\a=\sqrt{|\a^2|}$, $\b=\sqrt{|\b^2|}$.}
The case of positive coupling constants corresponds to the Snyder model on a spherical
background, while the model with negative $\a^2$ and $\b^2$ gives rise to the anti-Snyder
model on a pseudosphere. We call the latter anti-Snyder-de Sitter (aSdS) model.

The \pb (2.1) can be obtained from those of the flat Snyder model, with position
variables $\cX_i$ and momentum variables $\cP_i$ obeying
$$\{\cX_i,\cX_j\}=\b^2\,(\cX_i\cP_j-\cX_j\cP_i),\qquad\{\cP_i,\cP_j\}=0,\qquad\{\cX_i,\cP_j\}=
\d_{ij}+\b^2\,\cP_i\cP_j,\eqno(2.2)$$
by performing a linear unimodular but non-symplectic transformation of the phase space
coordinates,
$$x_i=\cX_i+{\b\over\a}\,\l\cP_i,\qquad p_i=(1-\l)\cP_i-{\a\over\b}\,\cX_i,\eqno(2.3)$$
where $\l$ is a free parameter, that can be chosen arbitrarily.
Actually, the freedom in the choice of a parameter $\l$ is already present in the flat
Snyder model. In that case, the symplectic \tran $\cX_i\to\cX_i+\l\cP_i$, $\cP_i\to\cP_i$
leaves the \pb (2.2) invariant. Analogously, the transformation from the representation
(2.3) of the SdS model with $\l=0$ to one with $\l\ne0$ is symplectic. The physical
results are therefore independent of the choice of $\l$.

In ref.\ [16] it has been shown that the \ps variables $\cX_i$ and $\cP_i$ can in turn be
written in terms of \coo $X_i$, $P_i$ that satisfy canonical Poisson brackets, by means of
the nonlinear transformation
$$\cP_i={P_i\over\sqrt{1-\b^2P_k^2}},\qquad\cX_i=\sqrt{1-\b^2P_k^2}\ X_i.\eqno(2.4)$$
Combining (2.3) and (2.4), the \coo $x_i$ and $p_i$ that satisfy the SdS \pb can be written
in terms of canonical \coo $X_i$, $P_i$, as
$$\eqalign{x_i=&\pres\ X_i+\l\,{\b\over\a}\,{P_i\over\pres},\cr
p_i=&-{\a\over\b}\pres\ X_i+(1-\l)\,{P_i\over\pres},}\eqno(2.5)$$
with inverse \tran
$$X_i={(1-\l)\a x_i-\l\b p_i\over\a\den},\qquad P_i={\b p_i+\a x_i\over\b\den}.\eqno(2.6)$$
It is important to note that if $\a^2,\b^2<0$, one must impose $|\a x_k+\b p_k|<1$.
The range of definition of the \coo depends therefore on the value of the momentum. A
possible
interpretation of this fact is that the radius of the pseudosphere is a function of the
momentum of the particle: this situation is common in DSR theories defined in curved spaces,
where the metric properties are momentum dependent [19,14].
\bigbreak

\section{2.2 Symmetries}
The full phase space is invariant under the $SO(5)$ or $SO(4,1)$ group, depending on the
sign of the coupling constants. However, from a physical point of view, the symmetries of
the configuration space are more interesting.
These are of course described by the groups $SO(4)$ or $SO(3,1)$, that leave invariant
a three-dimensional space of constant positive (resp.\ negative) curvature, and
are generated by the angular momentum $J_{ij}=x_ip_j-x_jp_i$ and the momentum $p_i$.

The phase-space \coo transform as vectors under the action of the generators of the
rotations $J_{ij}$,
$$\{J_{ij},x_k\}=\d_{ik}x_j-\d_{ij}x_k,\qquad
\{J_{ij},p_k\}=\d_{ik}p_j-\d_{ij}p_k,\eqno(2.7)$$
The symmetry under rotations is therefore realized in the usual way.

The action of the translation generators $p_i$ on phase-space variables can be read from
(2.1).
While the momenta transform according to the standard law for a constant curvature
space, the action of the translations on the position \coo is deformed and  is nonlinear.

Due to the Jacobi identities, the fundamental \pb (2.1) transform covariantly under the
action of the symmetry group.

\section{2.3 Classical motion}
The Hamiltonian for a free particle can be defined in the usual way as
$$H={p_i^2\over2m}.\eqno(2.8)$$
The Hamilton equations are obtained taking into account the deformed Poisson structure,
and read
$$\dot x_i=(1+\b^2p_k^2+2\a\b\,x_kp_k)\,p_i+\a^2x_kp_k\,x_i,\qquad
\dot p_i=-\a^2(p_k^2\,x_i-x_kp_k\, p_i).\eqno(2.9)$$
The relation between velocity and momentum is no longer linear, and the equations are
difficult to solve. Nevertheless, the solution can easily be obtained in one dimension.
In this case, the \pb reduce to
$$\{x,p\}=1+(\a x+\b p)^2,\eqno(2.10)$$
and the Hamilton equations read
$$\dot x=[1+(\a x+\b p)^2]{p\over m},\qquad\qquad\dot p=0.\eqno(2.11)$$
Hence the momentum $p$ is constant, $p=p_0$, and integration of the first equation yields
$$x={1\over\a}\tan {\a p_0\over m}\,t-{\b\over\a}\,p_0,\qquad{\rm if\ }\a^2>0,$$
$$x={1\over\a}\tanh {\a p_0\over m}\,t-{\b\over\a}\,p_0,\qquad{\rm if\ }\a^2<0.$$
Notice that in the last case $|\a x+\b p|<1$, as required.

\bigskip
A more interesting example of dynamics is the one-dimensional harmonic oscillator,
with Hamiltonian
$$H={p^2\over2m}+{m\o_0^2x^2\over2}.\eqno(2.12)$$

Let us first consider the case $\a^2,\b^2>0$.
For unit mass, the Hamilton equations read
$$\dot x=[1+(\a x+\b p)^2]\,p,\qquad\qquad\dot p=-\o_0^2[1+(\a x+\b p)^2]\,x.\eqno(2.13)$$
Using (2.13), it is easy to check that, like in standard classical mechanics, the Hamiltonian
is conserved, namely
$${p^2\over2}+{\o_0^2x^2\over2}=E,\eqno(2.14)$$
with $E$ the total energy of the oscillator.

The \eom can be solved as follows: define  $z=\arctan(\a x+\b p)$. From eqs.\ (2.13)
it follows that
$$\dot z=\a p-\o_0^2\b x.\eqno(2.15)$$
Deriving (2.15), and substituting (2.13), one obtains
$$\ddot z=-\o_0^2\,{\sin z\over\cos^3 z}\,,\eqno(2.16)$$
which admits the first integral
$$\ha\dot z^2+{\o_0^2\over2}\tan^2z=\o_0^2\cE,\eqno(2.17)$$
with $\cE$ an integration constant, proportional to $E$. These equations are identical to those
of a classical particle moving in the effective potential $V=\o_0^2\tan^2 z$, and have the same
analytical form as those obtained in [16] for the flat Snyder model.

Integration of (2.17) yields
$$z=\a x+\b p={\sqrt{2\cE}\ \sin\o t\over\sqrt{1+2\cE\cos^2\o t}},\eqno(2.18)$$
where $\o=\sqrt{1+2\cE}\ \o_0$.
From (2.15) it is then easy to obtain
$$\a p-\o_0^2\,\b x={{\o_0\sqrt{2\cE(1+2\cE)}\ \cos\o t\over
\sqrt{1+2\cE\cos^2\o t}}}.\eqno(2.19)$$

Substituting in (2.14), one finds the relation between $\cE$ and $E$,
$$\cE=\g E,\qquad{\rm with}\qquad \g=\b^2+{\a^2\over\o_0^2},\eqno(2.20)$$
and then
$$\o=\sqrt{1+2\g E}\ \o_0.\eqno(2.21)$$
The solution can finally be written as
$$\eqalign{x=&\sqrt{2E\over\g}\ {{\a\over\o_0}\sin\o t-\b\sqrt{1+2\g E}\,\cos\o t\over
\o_0\sqrt{1+2\g E\cos^2\o t}}\,,\cr
p=&\sqrt{2E\over\g}\ {\b\sin\o t+{\a\over\o_0}\sqrt{1+2\g E}\,\cos\o t\over
\sqrt{1+2\g E\cos^2\o t}}\,.}\eqno(2.22)$$
%\bigskip
\centerline{\epsfysize=5truecm\epsfbox{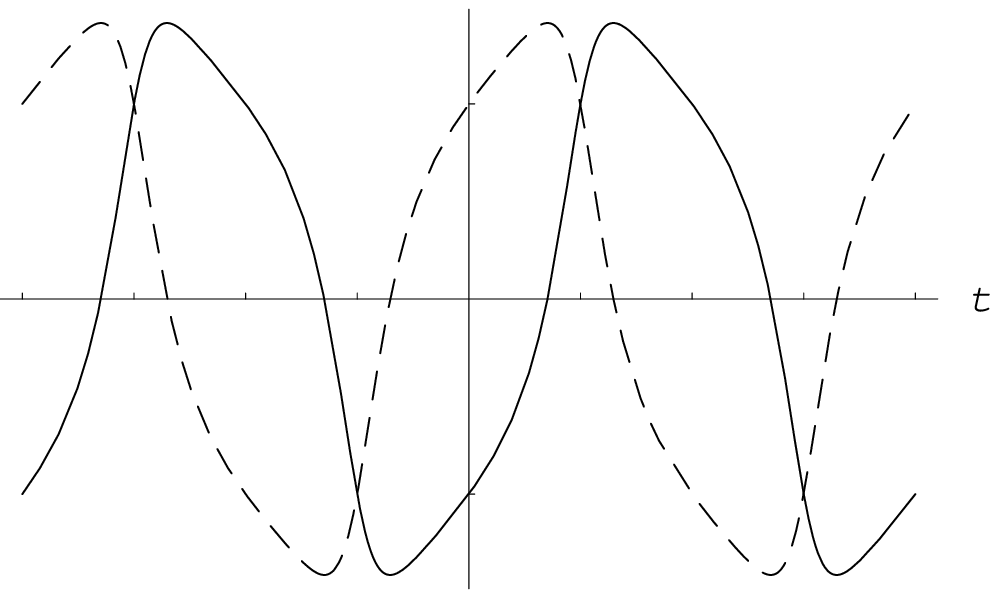}}

\medskip
{\baselineskip10pt
\centerline{\noindent\small Fig. 1: The solution of the SdS oscillator for
$\scriptstyle \g E=3/2$. The solid line}
\centerline{\quad\quad\ {\small represents the coordinate}
${\scriptstyle x}${\small, the dashed line the coordinate} {\smallmath p}{\small.}}}
\bigskip
\ni It appears that the solutions of the harmonic oscillator are periodic, but not
sinusoidal, see Fig.\ 1. Moreover, their frequency depends on the energy of the oscillator.
In order for the frequency to be real, the energy must be such that $1+2\g E\ge0$.
This is always true if the energy is positive. The flat Snyder oscillator [16] is
recovered in the limit $\a\to0$.

Let us now consider the aSdS model. The calculations are essentially identical to the
previous case, except that now one defines a variable $z=\arcth(\a x+\b p)$, that yields
an effective potential $V=\o_0^2\tanh z$. Proceeding as above, one recovers the relation
(2.21) between frequency and energy (but now with negative $\g$) and
$$\eqalign{x=&\sqrt{2E\over|\g|}\ {{\a\over\o_0}\sin\o t-\b\sqrt{1+2\g E}\,\cos\o t\over
\o_0\sqrt{1+2\g E\cos^2\o t}}\,,\cr
p=&\sqrt{2E\over|\g|}\ {\b\sin\o t+{\a\over\o_0}\sqrt{1+2\g E}\,\cos\o t\over
\sqrt{1+2\g E\cos^2\o t}}\,.}\eqno(2.23)$$
The character of the motion is similar to that found in the previous case, see Fig.\ 2.
Again, one must impose the condition $0<E<-1/2\g$ for the solutions to be real, that now
gives rise to an upper bound for the energy, $2E<\o_0^2/(|\b^2|\,\o_0^2+|\a^2|)$,
consistently with the bound $|\a x+\b p|<1$.
In the limit $\a\to0$, this condition does not depend on the frequency and is the same
that holds for the anti-Snyder model [16], while for $\b\to0$, gives $E<1/2|\b^2|$,
independently of $\o_0$.

%\bigskip
\centerline{\epsfysize=5truecm\epsfbox{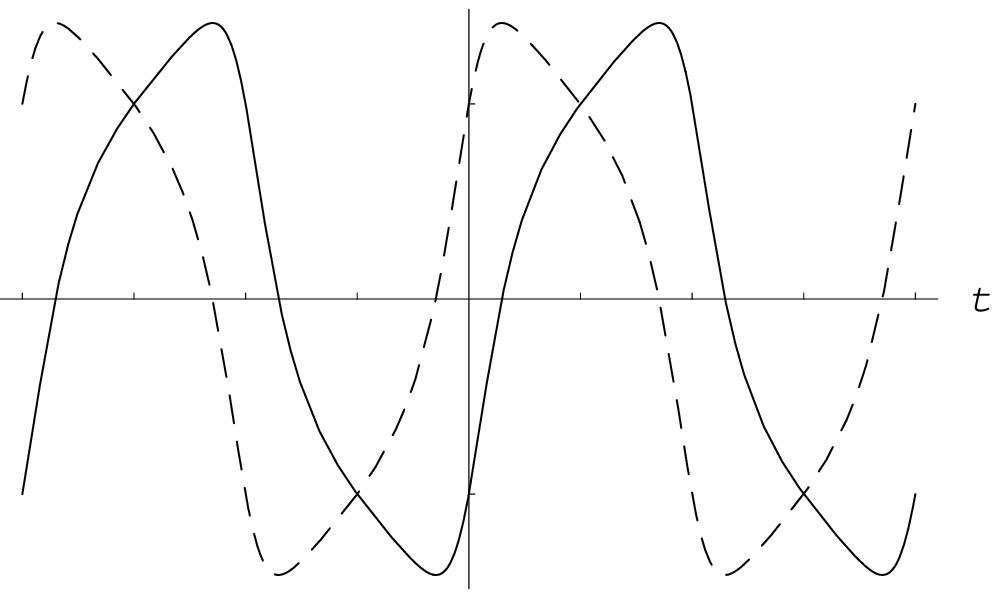}}
\medskip
{\baselineskip10pt
\centerline{\noindent\small Fig. 2: The solution of the aSdS oscillator for
$\scriptstyle \g E=-3/8$. The solid line}
\centerline{\quad\quad\ {\small represents the coordinate} ${\scriptstyle x}${\small,
the dashed line the coordinate} {\smallmath p}{\small.}}}
\bigbreak

\section{3. QUANTUM KINEMATICS}
The map (2.5) can be exploited to construct the quantum operators for position and momentum
that obey the SdS algebra. The deformation of the \cor implies a modification of the Heisenberg
uncertainty relations and gives rise to minimal uncertainty for both position and momentum. The
spectrum of the corresponding operators is changed accordingly.
Modified uncertainty relations have been studied in different contexts in several papers [3,18].

\section{3.1 Commutation relations and uncertainty}

For the (a)SdS model, the \cor (1.2) between the position operators $\hat x_i$ and the momentum
operators $\hat p_i$ can be written in the equivalent form
$$[\hx_i,\hx_j]=i\hbar\b^2\hat J_{ij},\qquad[\hp_i,\hp_j]=i\hbar\a^2\hat J_{ij},$$
$$[\hx_i,\hp_j]=i\hbar\[\d_{ij}+\a^2\hx_i\hx_j+\b^2\hp_i\hp_j+
{\a\b\over2}(3\hp_i\hx_j+\hx_j\hp_i-\hp_j\hx_i+\hx_i\hp_j)\],
\eqno(3.1)$$
where the last equation is obtained from (1.2) by substiting the representation $\hat J_{ij}=
\ha(\hp_i\hx_j+\hx_j\hp_i-\hp_j\hx_i-\hx_i\hp_j)$ for the angular momentum. The lower degree of
symmetry of (3.1) \wrt the classical expression (2.1) is due to operator ordering problems.

In the simple case in which $\brah p_i\ket=\brah x_i\ket=0$, the \ur that follow from (1.2)
for states with vanishing angular momentum are
$$\eqalign{\D x_i\D p_j&\ge{\hbar\over2}\left|\bra[\hx_i,\hp_j]\ket\right|={\hbar\over2}
\,\left|\d_{ij}+\bra(\a\hx_i+\b\hp_i)(\a\hx_j+\b\hp_j)\ket\right|\cr
&\ge{\hbar\over2}\,\left|\d_{ij}+\a^2\D x_i\D x_j+\b^2\D p_i\D p_j-\a\b(\D x_i\D p_j+
\D x_j\D p_i)\right|,}\eqno(3.2)$$
where the Schwartz inequality has been used in the last step.
In the one-dimensional case, the algebra (3.1) simplifies greatly. In particular,
$$[\hx,\hp]=i\hbar\[1+\a^2\hx^2+\b^2\hp^2+\a\b(\hx\hp+\hp\hx)\],\eqno(3.3)$$
and the \ur (3.2) reduce to
$$\D x\D p\ge{\hbar\over2}\,{\left|1+\a^2(\D x)^2+\b^2(\D p)^2\right|\over1+\hbar\a\b}.
\eqno(3.4)$$
If $\a^2,\b^2>0$, they imply the existence of both  minimal position and
momentum uncertainties, given by
$$\D x_M={\hbar\b\over\sqrt{1+2\hbar\a\b}}\sim\hbar\b(1-\hbar\a\b),\qquad
\D p_M={\hbar\a\over\sqrt{1+2\hbar\a\b}}\sim\hbar\a(1-\hbar\a\b).
\eqno(3.5)$$
The limit values may not be achieved by physical states.
With the choice of parameters discussed in sect.\ 1, the numerical value of the
minimal uncertainties are the Planck length, $\D x_M\sim10^{-33}$cm, and an extremely tiny
momentum, $\D p_M\sim10^{-62}$g.

No minimal uncertainties emerge instead if $\a^2,\b^2<0$.

\section{3.2 Representation of position and momentum operators}
Extending the transformations (2.5) to the quantum case, it is possible to define a realization
of the position and momentum operators $\hat x_i$ and $\hat p_i$ that obeys the algebra (3.1).
More precisely, the operators act on a Hilbert space of functions of a variable $P_i$, as
$$\eqalign{\hat x_i&=\hat\cX_i+\l\,{\b\over\a}\,\hat\cP_i=i\pres\ {\de\over\de P_i}+
\l\,{\b\over\a}\,{P_i\over\pres},\cr
\hat p_i&=-{\a\over\b}\hat\cX_i+(1-\l)\hat\cP_i=-i{\a\over\b}\pres\ {\de\over\de P_i}+
(1-\l)\,{P_i\over\pres},}\eqno(3.6)$$

If $\a^2,\b^2>0$, the range of allowed values of $P_i$ is bounded by $P_k^2<1/\b^2$.
If $\a^2,\b^2<0$, instead, all real values of $P_i$ are allowed, but the upper bound of the
eigenvalues of the operator $(\a\hx_k+\b\hp_k)^2$ is 1, as in the classical limit.

In the previous formulae, the realizations
corresponding to different values of $\l$ are, at least formally, unitary equivalent.
For example, in the case $\a^2,\b^2>0$, they are related through the action of the operator $U=(1-\b^2P_k^2)^{i\l\over2\hbar\a\b}$,
so that $\hx_i(\l)=U^\dagger\hx_i(0)U$, \hbox{$\hp_i(\l)=U^\dagger\hp_i(0)U$}.
One is therefore free to choose the value of $\l$ without affecting the physical
predictions of the theory, and hence in the following, unless explicitly stated, we shall
set $\l$ to 0. An exception will be the investigation of the harmonic oscillator, where a
different choice is more convenient.

\bigskip

We now discuss in detail the action of the operators in the simple case of a particle in
one dimension in the representation (3.6) with $\l=0$.
When \hbox{$\a^2,\b^2>0$}, in order to get symmetric operators, i.e.
$$(\hat p\,\q,\f)=(\q,\hat p\,\f),\qquad(\hat x\,\q,\f)=(\q,\hat x\,\f),\eqno(3.7)$$
the scalar product must be defined as
$$(\q,\f)=\int_{-1/\b}^{1/\b}{dP\over\sqrt{1-\b^2P^2}}\ \q^*(P)\,\f\,(P),\eqno(3.8)$$
and the wave functions must satisfy periodic boundary conditions, $\q(-1/\b)=\q(1/\b)$.
This is readily checked like in ref.\ [16].

If instead $\a^2,\b^2<0$, the scalar product is defined as
$$(\q,\f)=\int_{-\inf}^{\inf}{dP\over\sqrt{1-\b^2P^2}}\ \q^*(P)\,\f\,(P),\eqno(3.9)$$
and only the convergence of the integral at infinity is required.

The study of the spectrum of the position and momentum operators encounters some
difficulties. For example, the eigenvalue equation
$$\hat x\,\q_x=x\,\q_x,\eqno(3.10)$$
has solutions
$$\q_x=C\,{\rm e}^{-{ix\over\hbar\b}\arcsin\b P},
\eqno(3.11)$$
with $C$ a normalization constant and $x=2n\hbar\b$.
The eigenstates (3.11) have vanishing position uncertainty, but their energy diverges,
so that they cannot be accepted as physical.
The same situation occurs in ordinary quantum mechanics, but in that case the position
eigenstates can be obtained as limit of states with finite energy. This is not possible
in our case, due to the existence of a minimal indetermination for the positions, which
implies that no exactly localized states can exist.
As in similar cases [18], one may try to define states with
minimal position uncertainty by suitably modifying the equation that defines the
position eigenstates, in order to take into account the finite uncertainty in position.
However, we were not able to find a suitable modification. Nevertheless, since
$[\hx,\hp]=[\hat{\cal X},\hat{\cal P}]$
one may guess that the functions that minimize the commutator \ev are the same as those
that give rise to minimal uncertainty in flat Snyder space, i.e.
$$\q_x=C\sqrt{1-\b^2P^2}\ \ex^{-{ix\over\hbar\b}\,\arcsin\b P}.\eqno(3.12)$$
Notice that the functions (3.12) satisfy the boundary conditions without need of
quantizing the parameter $x$. For these functions,
$$\D x=\hbar\b,\qquad\D p=\sqrt{{1\over\b^2}+\hbar^2\a^2},
\qquad\D x\D p=\hbar\,\sqrt{1+(\hbar\a\b)^2},\eqno(3.13)$$
and $\D x$ is close to its minimal value (3.5), while $\D p$ is extremely
large, due to the smallness of $\b$.

More general functions in the domain of $\hx$ and $\hp$ are given by
$$\q_x=C(1-\b^2P^2)^\r\ \ \ex^{-{ix\over\hbar\b}\,\arcsin\b P},\eqno(3.14)$$
with ${\rm Re}\; \r>{1\over4}$, but these give rise to greater uncertainties.
For example, for $\r=\ha+i\m$, one has
$$(\D x)^2=\hbar^2\b^2+{\m^2\over\a^2},\qquad(\D p)^2=\hbar^2\a^2+{(1-\m)^2\over\b^2}.
\eqno(3.15)$$
We notice that the value of $\D x\D p$ can be minimized for this class of functions
by the choice $\m=(1\pm\sqrt{1-4\hbar^2\a^2\b^2})/2$, for which $\D x\D p=\hbar$.

\medskip
The properties of the momentum operator $\hp$ are similar to those of $\hx$.
The eigenvalue equation
$$\hat p\,\q_p=p\,\q_p,\eqno(3.16)$$
has solution
$$\q_p=C(1-\b^2P^2)^{i\over2\hbar\a\b}\ {\rm e}^{{ip\over\hbar\a}\arcsin\b P},
\eqno(3.17)$$
with $p=2n\hbar\a$, and presents problems analogous to those affecting the eigenfunctions
of the position operator. In this case $\D p=0$, but the expectation value of $\hx^2$
diverges. As before, however, one can define a basis, analogous to (3.12),
$$\q_p=C(1-\b^2P^2)^{\ha+{i\over2\hbar\a\b}}\  \ex^{{ip\over\hbar\a}\,\arcsin\b P},
\eqno(3.18)$$
where $p$ is not quantized, for which the expectation value of $\hx^2$ is finite.
The uncertainties in this basis are
$$\D x=\sqrt{{1\over\a^2}+\hbar^2\b^2},\qquad\D p=\hbar\a,
\qquad\D x\D p=\hbar\,\sqrt{1+(\hbar\a\b)^2}.\eqno(3.19)$$
More general functions analogous to (3.14) may also be considered.

To summarize, although formal eigenstates (3.11), (3.17) give rise to a discrete spectrum
for both position and momentum, they are not physical because have infinite \ev of
$\hp^2$ or $\hx^2$.
A more physical basis of states is given by smeared functions like (3.14), (3.18), that
describe a fuzzy phase space, with no sharply defined values of positions and momenta.
\bigskip
In the aSdS case, the situation is instead analogous to ordinary quantum mechanics, since
no minimal indetermination arises, and the formal eigenfunctions are limit of
states with finite $\D x$ and $\D p$.
For the position operator, the eigenfunction are
$$\q_x=C\,\ex^{-{ix\over\hbar\b}\,\arcsh\b P},\eqno(3.20)$$
and for momentum,
$$\q_p=C(1-\b^2P^2)^{i\over2\hbar\a\b}\ {\rm e}^{{ip\over\hbar\a}\arcsh\b P}.
\eqno(3.21)$$

\section{3.3 Quantum symmetries}
The invariance of the configuration space of the classical model under the $SO(4)$ or $SO(3,1)$
group can be extended to the quantum case.

The rotations are generated by
$$\hat J_{ij}=\ha(\hat x_i\hat p_j+\hat p_j\hat x_i-\hat x_j\hat p_i-\hat p_i\hat x_j)=
\hat\cP_j\hat\cX_i-\hat\cP_i\hat\cX_j=i\hbar\left(P_j{\de\over\de P_i}-P_i{\de\over\de P_j}
\right).\eqno(3.22)$$
and act in the standard way. In particular, the spectrum of $\hat J_{ij}$ is the same as in
ordinary quantum mechanics. Defining $\hat L_i=\e_{ijk}\hat J_{jk}$, the eigenfunctions in spherical
coordinates in momentum representation are given by
the standard spherical harmonics
$$\hat L^2\,Y_{lm}(P_\h,P_\f)=l(l+1)\,Y_{lm}(P_\h,P_\f),\qquad
\hat L_z\,Y_{lm}(P_\h,P_\f)=m\,Y_{lm}(P_\h,P_\f).\eqno(3.23)$$

The translations are generated by the momentum operators $\hat p_i$, that act according to (3.1).
While the action of the translations on momenta is the usual one for a space of constant curvature,
that on position variables is deformed and takes a nonlinear form.
The \cor (3.1) transform covariantly under these symmetries.

\section{4. QUANTUM DYNAMICS IN ONE DIMENSION}
We shall now use the results of the previous section to study the solutions of some simple
one-dimensional quantum systems. We write down in detail the calculations for the SdS model,
and simply report the results for the aSdS case, that can be investigated in the same way.

\section{4.1. Free particle}
Let us first consider the \schr equation for a free particle in one dimension.
In the \rep (3.6) with $\l=0$, for unit mass it reads
$${d^2\q\over dP^2}-\left(\b-{2i\over\hbar\a}\right){\b P\over1-\b^2P^2}\ {d\q\over dP}-
{\b^2\over\hbar^2\a^2}\left[{P^2-i\hbar\a/\b\over(1-\b^2P^2)^2}-{2E\over1-\b^2P^2}\right]\q=0.
\eqno(4.1)$$
In the SdS case, solutions of (4.1) that vanish at $P=\pm1/\b$ exist, of the form
$$\q=\const.\times(1-\b^2P^2)^{i\over2\hbar\a\b}\ \cos\[{\sqrt{2E}\over\hbar\a}\arcsin\b P\],
\eqno(4.2)$$
with $E=\ha\hbar^2\a^2n^2$, for odd integer $n$. These solutions have finite values
of $\D x$.

For aSdS, the relevant solutions are instead given by the momentum eigenfunctions (3.21),
and the energy is not quantized.

\section{4.2. Harmonic oscillator}
Let us now consider the one-dimensional quantum harmonic oscillator, with Hamiltonian
$$H={\hat p^2\over2m}+{m\o_0^2\hat x^2\over2}.\eqno(4.3)$$
In order to simplify the calculations, we exploit the possibility of choosing the coefficient
$\l$ in the representation (3.6) so that the cross terms $\hat\cP\hat\cX+\hat\cX\hat\cP$ in
the Hamiltonian vanish, setting
$$\l={\a^2\over\b^2\o_0^2+\a^2}.\eqno(4.4)$$

With this choice, the \schr equation reads, for unit mass,
$$\ha\,{\b^2\o_0^2\over\b^2\o_0^2+\a^2}\left[\hat\cP^2+{(\b^2\o_0^2+\a^2)^2\over\b^4\o_0^2}\
\hat\cX^2\right]\q=E\q.\eqno(4.5)$$
Using the realization (3.6) of the operators, eq.\ (4.5) can be written as
$${d^2\q\over dP^2}-{\b^2P\over1-\b^2P^2}\ {d\q\over dP}-{1\over\hbar^2\o^2}
\left[{P^2\over(1-\b^2P^2)^2}-{2{\cal E}\over1-\b^2P^2}\right]\q=0,\eqno(4.6)$$
where $\o=\left(1+{\a^2\over\b^2\o_0^2}\right)\o_0$ and
${\cal E}=\left(1+{\a^2\over\b^2\o_0^2}\right)E$.

The form of this equation is the same that holds for the flat Snyder model [16], but with
different coefficients, and can be solved in the same way. In particular,
when $\a^2,\b^2>0$,
one can define a variable $\bar P=\arcsin\b P$, in terms of which (4.6) becomes the
standard \schr equation for a potential
$$V={1\over\o^2}\ \tan^2\bar P,\eqno(4.7)$$
that coincides with the classical potential of sect.\ 2.

In order to find the explicit solution of eq.\ (4.6), it is more convenient, however, to
define the variable $z=(1+\b P)/2$, in terms of which the equation can be written in the
hypergeometric form
$${d^2\q\over dz^2}+{z-\ha\over z(z-1)}\ {d\q\over dz}-
\left[{\m(z-\ha)^2\over z^2(z-1)^2}+{\e\over z(z-1)}\right]\q=0,\eqno(4.8)$$
with
$$\m={\o_0^2\over\hbar^2(\b^2\o_0^2+\a^2)^2},\qquad\e={2E\over\hbar^2(\b^2\o_0^2+\a^2)}.$$
By standard methods, one can then obtain the solution in terms of the hypergeometric
function ${\bf F}(a,b,c;z)$,
$$\q=\const\times(1-\b^2P^2)^{(1+\sqrt{1+4\m})/4}\ {\bf F}\left(a,b,c;{1+\b P\over2}\right),
\eqno(4.9)$$
where
$$a=\ha(1+\sqrt{1+4\m})-\sqrt{\m+\e},\quad b=\ha(1+\sqrt{1+4\m})+\sqrt{\m+\e},\quad
c=1+\ha\sqrt{1+4\m}\,.$$

We require that $\q$ vanish at $P=\pm1/\b$, i.e.\ at $z=0,1$. This occurs when $a=-n$ or
$b=-n$. In both cases,
$$\e=\left(n+\ha\right)(1+\sqrt{1+4\m})+n^2,\eqno(4.10)$$
and the solution can be written in terms of Gegenbauer polynomials ${\bf C}^\a_n$ [20] as
$$\q=\const\times(1-\b^2P^2)^{\a/2}{\bf C}^\a_n(\b P),\eqno(4.11)$$
with $\a=\ha\(1+\sqrt{1+4\m}\)$.

From (4.10) follows the spectrum of the energy,
$$E=\left(n+\ha\right)\hbar\o_0\sqrt{1+{\hbar^2(\b^2\o_0^2+\a^2)^2\over4\o_0^2}}+
\left(n^2+n+\ha\right){\hbar^2(\b^2\o_0^2+\a^2)\over2},\eqno(4.12)$$
which exhibits corrections of order $\hbar\,(\b^2\o_0+\a^2/\o_0)$ \wrt the standard case
and a duality for $\b^2\o_0\lra\a^2/\o_0$.

In the limit $\a\to0$ one recovers the results of [16] for flat Snyder space, whereas for
$\b\to0$, one obtains the energy spectrum on a 3-sphere,
$$E=\left(n+\ha\right)\hbar\o_0\sqrt{1+{\hbar^2\a^4\over4\o_0^2}}+
\left(n^2+n+\ha\right){\hbar^2\a^2\over2},\eqno(4.13)$$
for which at first order the shift in the energy \wrt the standard oscillator is independent
of $\o_0$.

\bigskip
The same calculation can be performed when $\a^2,\b^2<0$. The energy spectrum is simply the
analytic continuation of (4.12) for negative values of $\a^2$ and $\b^2$. In this case, for
great $n$ the energy becomes negative. In order to ensure positivity of energy, one must
therefore impose an upper bound on the allowed values of $n$.

\section{5. QUANTUM DYNAMICS IN THREE DIMENSIONS}

Of course, the extension of the previous results to three-dimensional space is of great physical
relevance, since it permits to test the effect of the noncommutativity of spatial coordinates.

The analysis of the previous sections easily generalizes to three dimensions.
Since the components of the position and momentum operators $\hat x_i$ and $\hat p_i$ do not
commute, we shall limit our consideration to spherically symmetric problems, where one can use
polar coordinates, that much simplify the problem.
Moreover, we set $\hbar=1$ in this section.

\section{5.1 Position eigenstates in three dimensions}

As in standard quantum mechanics, a basis of operators for one-particle states in
momentum \rep is given by the radial component of the momentum $\hp_r$
and by the angular momentum $\hat L_i$, introduced in section 3.3.

Therefore, we adapt the representation (3.6) to three-dimensional spherical coordinates
in the $P_i$-space. We first define the operators
$$\hat\cP_r=\cP_r\id\sqrt{\cP_i^2}=\sqrt{P_i^2\over1-\b^2P_i^2},\qquad
\hat\cX_r=\sqrt{1-\b^2P_r^2}\(i{\de\over\de P_r}+{1\over P_r}\),\eqno(5.1)$$
with $[\hat\cX_r,\hat\cP_r]=i$. As in ordinary \QM it follows that
$$\hat\cP_i^2=\hat\cP_r^2,\qquad\hat\cX_i^2=\hat\cX_r^2+{\hat L^2\over\cP_r^2},\qquad
\hat\cX_i\hat\cP_i+\hat\cP_i\hat\cX_i=\hat\cX_r\hat\cP_r+\hat\cP_r\hat\cX_r,\eqno(5.2)$$
where $\hat L^2$ is the square of the angular momentum operator. The square of the position
and momentum operators (3.6) can then be written in terms of the radial operators as
$$\hat x_i^2=\(\hat\cX_r+{\b\over\a}\,\l\,\hat\cP_r\)^2+{\hat L^2\over\cP_r^2},
\qquad\hat p_i^2=\((1-\l)\hat\cP_r-{\a\over\b}\,\hat\cX_r\)^2+{\a^2\over\b^2}
{\hat L^2\over\cP_r^2}.\eqno(5.3)$$

Since the action of the  angular momentum is the standard one, as discussed in section 3.3,
a generic three-dimensional wave function can be expanded in spherical harmonics, as
defined in (3.23),
$$\q(P_r,P_\h,P_\f)=\sum_{l,m}\q_{rlm}(P_r)\, Y_{lm}(P_\h,P_\f),\eqno(5.4)$$
and only the radial functions need to be investigated in detail.
In the following we shall omit the $lm$ indices in the radial functions.

The scalar product in the space of the radial functions can be defined as
$$(\q_r,\f_r)=\int_0^{1/\b}{P_r^2\,dP_r\over\prep}\ \q_r^*(P_r)\,\f_r\,(P_r).\eqno(5.5)$$

The spectrum of the radial momentum and position operators will be analogous to the
one of  the corresponding one-dimensional operators, except that now $P_r$ can only take
positive values.  Therefore we shall not discuss it in detail, but rather pass to investigate
the \schr equation.

The basis adopted in this section also permits to immediately single out the states that
minimize the \ur between the position coordinates along different directions. In fact,
from (3.1) it follows that
$$\D x_i\D x_j\ge\left|{\b^2\over2}\ \bra\hat J_{ij}\ket\right|,\eqno(5.6)$$
and the states that minimize these \ur are those with vanishing angular momentum.
The same considerations hold for the components of the momentum.

The previous discussion can easily be extended to the anti-Snyder model.

\section{5.2 The free particle in three dimensions}
Let us  consider the \schr equation for a free particle in three dimensions.
Choosing the gauge $\l=0$, and using the \rep (5.1), (5.3) and the expansion (5.4)
in spherical harmonics, the radial part of the \schr equation can be written as
$$\eqalign{{d^2\q_r\over dP_r^2}&-{\left(3\b^2-2i{\b\over\a}\right)P_r^2-2\over
P_r(1-\b^2P_r^2)}\ {d\q_r\over dP_r}\cr
&-{\b^2\over\a^2}\left[{(1-4i\a\b)P_r^2+3i{\a\over\b}\over(1-\b^2P_r^2)^2}
+{l(l+1)\a^2\over\b^2P_r^2}-{2E\over1-\b^2P_r^2}\right]\q_r=0.}
\eqno(5.7)$$
Defining now a function $u(P_r)$ such that $\q_r=(1-\b^2P_r^2)^{i/2\a\b}u$,
eq.\ (5.7) simplifies to
$${d^2u\over dP_r^2}+\({2\over P_r}-{\b^2P_r\over1-\b^2P_r^2}\){du\over dP_r}
-\[{l(l+1)\over P_r^2}-{2\b^2E\over \a^2(1-\b^2P_r^2)}\]u=0.
\eqno(5.8)$$

Finally, after a change of variable $z=\b^2P_r^2$, the equation takes the form of a
\hyp differential equation,
$${d^2u\over dz^2}+{3-4z\over2z(1-z)}{du\over dz}-{1\over4}\[{l(l+1)\over z^2}-
{\e\over z(1-z)}\]u=0,\eqno(5.9)$$
where $\e=2E/\a^2$, with solution
$$u(P_r)=\const\times\sqrt{1-\b^2P_r^2}\;(\b P_r)^l\,{\bf F}\left(a,b,c;\,\b^2P_r^2\right),
\eqno(5.10)$$
with ${\bf F}$ a hypergeometric function of parameters
$$a=1+{l\over2}+{\sqrt{1+\e+l(l+1)}\over2},\quad
b=1+{l\over2}-{\sqrt{1+\e+l(l+1)}\over2},\quad
c=l+{3\over2}\,.$$
The solution of (5.7) is therefore
$$\q_r=(1-\b^2P_r^2)^{\ha(1+i/\a\b)}(\b P_r)^l\,{\bf F}\left(a,b,c;\,\b^2P_r^2\right).
\eqno(5.11)$$
The \bc must be fixed in such a way that $\q_r$ vanish at $P_r=1/\b$, i.e.\ at $z=1$.
This occurs when $b=-n$, with integer $n$. It follows that
$$E=\a^2\(2n^2+4n+2nl+{3\over2}\,l+{3\over2}\).$$
Hence, in the SdS case, the eigenvalues of the energy are quantized. This is of course a
consequence of the finite range of the coordinate $P_r$.
The radial wave function can be written
$$\q_r=\const.\times(1-\b^2P_r^2)^{\ha(1+i/\a\b)}\,(\b P_r)^l\ \tP_n^{(l+\ha,\ha)}
(1-2\b^2P_r^2),\eqno(5.12)$$
with $\tP_n^{(\m,\n)}$ a Jacobi polynomial [20].

For the spherical wave, $l=0$, the solution (5.12) takes a simple form. Using the
properties of the \hyp functions, one easily gets
$$\q_{r0}=(1-\b^2P_r^2)^{i/2\a\b}\ {\sin\[\sqrt{1+\e}\,\arcsin \b P_r\]
\over\sqrt{1+\e}\ \b P_r},$$
that in the limit $\a\to0$, $\b\to0$ coincides with the standard quantum-mechanical
solution $\q_{r0}=\sin(\sqrt{2E}\,P_r)/(\sqrt{2E}\,P_r)$.

In the aSdS case, the \schr equation is the analytic continuation of (5.7) to negative
values of $\a^2$ and  $\b^2$. Requiring regularity for $P_r\to\inf$, the solution reads
$$\q_r=\const.\times {(1-\b^2P_r^2)^{i/2\a\b}\over\(\b^2P_r^2\)^a}\
{\bf F}\(a,a-c+1,a-b+1,{1\over\b^2P_r^2}\)\eqno(5.13)$$
and does not imply a quantization of the energy.

\section{5.3 The harmonic oscillator in three dimensions}
In order to write the \schr equation for the 3-dimensional harmonic oscillator, one can proceed
as for the free particle. However, in this case it is convenient to choose $\l$ as in (4.4),
in order to remove the mixed terms from the equation. With this choice, one obtains
$$\ha\,{\b^2\o_0^2\over\b^2\o_0^2+\a^2}\left[\hat\cP_i^2+{(\b^2\o_0^2+\a^2)^2\over\b^4\o_0^2}\
\hat\cX_i^2\right]\q=E\q.\eqno(5.14)$$
Expanding in spherical harmonics as in (5.4), after some algebraic manipulations the equation
for the radial wave function becomes
$$\eqalign{{d^2\q_r\over dP_r^2}&+\({2\over P_r}-{\b^2P_r\over1-\b^2P_r^2}\){d\q_r\over dP_r}\cr
&-\[{l(l+1)\over P_r^2}+{1\over\o^2}{P_r^2\over(1-\b^2P_r^2)^2}-
\({2\cE\over\o^2}-\b^2\){1\over1-\b^2P_r^2}\]\q_r=0,}\eqno(5.15)$$
where $\o=\left(1+{\a^2\over\b^2\o_0^2}\right)\o_0$ and
${\cal E}=\left(1+{\a^2\over\b^2\o_0^2}\right)E$.

Defining a new variable $z=\b^2P_r^2$, (5.15) can be put in the form of a hypergeometric
equation,
$${d^2\q_r\over dz^2}+{3-4z\over2z(1-z)}{d\q_r\over dz}-{1\over4}\[{l(l+1)\over z^2}+
{\m\over(1-z)^2}-{\e\over z(1-z)}\]\q_r=0,\eqno(5.16)$$
with
$$\m={1\over\b^4\o^2},\qquad\e={2\cE\over\b^2\o^2}-1.$$
Note that (5.16) differs from the free particle equation (5.9) only for the term proportional
to $\m$. The solution of (5.16) can be written as
$$\q_r=\const\times(\b P_r)^l(1-\b^2P_r^2)^{(1+\sqrt{1+4\m})/4}\ {\bf F}\left(a,b,c;\,
\b^2P_r^2\right),\eqno(5.17)$$
with
$$\eqalign{&a=\ha\(l+{3\over2}+{\sqrt{1+4\m}\over2}+\sqrt{1+\m+\e+l(l+1)}\),\cr
&b=\ha\(l+{3\over2}+{\sqrt{1+4\m}\over2}-\sqrt{1+\m+\e+l(l+1)}\),\cr
&c=l+{3\over2}\,.}$$

We require that $\q$ vanish at $P_r=1/\b$, i.e.\ at $z=1$. This occurs when
$b=-n$, and then
$$\e=\(2n+l+{3\over2}\)\sqrt{1+4\m}+4n^2+4nl+6n+2l+{3\over2},\eqno(5.18)$$
and
$$\q_r=\const\times(\b P_r)^l(1-\b^2P_r^2)^{(1+\sqrt{1+4\m})/4}\
\tP_n^{(l+\ha,{\sqrt{1+4\m}\over2})}(1-2\b^2P_r^2),\eqno(5.19)$$
with $\tP_n^{(\m,\n)}$ a Jacobi polynomial.
From (5.18) it follows that
$$E=\left(2n+l+{3\over2}\right)\o_0\sqrt{1+{(\b^2\o_0^2+\a^2)^2\over4\o_0^2}}+
\left(2n^2+3n+2nl+l+{5\over4}\right)(\b^2\o_0^2+\a^2).\eqno(5.20)$$
The previous expression may also be written in terms of a new quantum number $N=2n+l$,
that is commonly introduced for the three-dimensional oscillator,
$$E=\left(N+{3\over2}\right)\o_0\sqrt{1+{(\b^2\o_0^2+\a^2)^2\over4\o_0^2}}+
\left(N^2+3N-l(l+1)+{5\over2}\right){\b^2\o_0^2+\a^2\over2}.\eqno(5.21)$$
Like in one dimension, corrections of order $(\b^2\o_0+\a^2/\o_0)$ \wrt ordinary \QM
occur in the spectrum of the harmonic oscillator.
In the limit $\a\to0$ one recovers the result valid for flat Snyder space, whereas
for $\b\to0$, one obtains the spectrum of the oscillator on a 3-sphere. In this case, at
first order the shift in the energy \wrt the standard oscillator is independent of $\o_0$.
\bigskip
The energy spectrum of the 3-dimensional aSdS harmonic oscillator can be
obtained by analytic continuation of that of the SdS oscillator to negative $\a^2$ and
$\b^2$, and hence is still given by (5.20).
As in one dimension, one must impose some conditions on the quantum numbers in
order to preserve the positivity of the energy.

\section{6. CONCLUSIONS}
We have studied the classical and the quantum mechanics of the free particle and of the harmonic
oscillator in the nonrelativistic version of the Snyder model on a constant curvature background.
The calculations have been based on the existence of a linear mapping
relating the (a)SdS phase space variables to flat Snyder coordinates.

As in the flat case, the physics strongly depends on the sign of the coupling constants in the
defining \pb (or \cor in the quantum case). In particular, we have been able to construct
the Snyder model on a space of constant positive curvature, and the anti-Snyder model on a
space of constant negative curvature. In both cases, the curvature of momentum space has the
same sign as that of position space.

In the SdS case, a bound on the localization of particles in both position and momentum space
arises, while this does not occur in aSdS.
Moreover, contrary to the flat anti-Snyder model, the range of values of the momentum in aSdS
is not limited,
although an upper bound holds for a linear combination of the position and momentum of a particle.
This can be interpreted as an example of momentum-dependent geometry, like in other
generalizations of DSR to curved spaces [19,14].
Even if we have not discusseed this topic, one can also employ the same arguments used for a flat
background [16], to show that quantization of area arises for SdS, but not for aSdS.

In the limit $\a\to0$, one recovers the results valid for flat Snyder space, while for $\b\to0$,
one obtains the
dynamics on a space of constant curvature. The study of this limit may also be useful
in the relativistic case for the investigation of motion in \ades spacetime.

The classical dynamics of free particles does not give rise to particularly relevant effects,
while in the quantum case the energy spectrum is discrete for SdS and continuous for aSdS.
The behavior of the one-dimensional harmonic oscillator is more interesting: both in the
classical and in the quantum cases, its frequency depends on the energy. Moreover, in aSdS
the quantum energy spectrum has only a finite number of eigenvalues.

The three-dimensional dynamics is similar to that governing the one-dimensional case, and can
be investigated by means of an expansion in spherical harmonics, exploiting the fact that the
angular momentum operator in (a)SdS enjoys standard properties. Its interest relies on the fact
that the effects of the noncommutativity of spacetime and momentum space can be displayed
explicitly. It turns out that minimal indeterminations are obtained for states with vanishing
angular momentum.

Another relevant result is obtained in the limit $\b\to0$, that, as mentioned above, gives the
dynamics on a 3-sphere.
To our knowledge, no exact solutions of the quantum harmonic oscillator on a 3-sphere have
previously been discussed in the literature.

It would be interesting to extend our results to more general systems, like for example the
Coulomb potential. However, when the potential is nonpolynomial problems arise because of the
nonlocality of the Hamiltonian when our transformation is applied, and only approximate
solutions can be found.

An alternative approach to the problem studied in this paper, analogous to that adopted in [17]
for the Snyder model,
would consist in the investigation of the representations of the symmetry group $SO(5)$ on the
coset space $SO(5)/(SO(3)\times SO(2))$, which is the phase space of three-dimensional Euclidean
SdS, or of the group $SO(4,1)$ on $SO(4,1)/(SO(2,1)\times SO(2))$ for aSdS.

Our study is limited to a nonrelativistic curved background. However, it can be used as a
starting point to extend our results to the relativistic case and to quantum field theory.
This would be important in the context of quantum gravity, where noncommutative curved
spacetimes are expected to play a crucial role. This topic is presently under investigation.

\section{Acknowledgements}
{\ni I wish to thank M. Stetsko for pointing out an error in a previous version of this paper.}
\vfill\eject
\beginref
\ref [1] H.S. Snyder, \PR{71}, 38 (1947).
\ref [2] S. Majid and H. Ruegg, \PL{B329}, 189 (1994);
 S. Doplicher, K. Fredenhagen and J.E. Roberts, \PL{B331}, 39 (1994).
\ref [3]  A. Kempf, \LMP{26}, 1 (1992);
 M. Maggiore, \PL{B304}, 63 (1993).
\ref [4] T. Padmanhaban, \AoP{165}, 38 (1985);
 G. Veneziano, Europhys. Lett. {\bf2}, 199 (1986).
\ref [5] C.A. Mead, \PR{135}, B849 (1964).
\ref [6] G. Amelino-Camelia, \PL{B510}, 255 (2001), \IJMP{D11}, 35 (2002);
J. Magueijo and L. Smolin, \PRL{88}, 190403 (2002).
\ref [7] J. Kowalski-Glikman and S. Nowak, \IJMP{D13}, 299 (2003).
\ref [8] S. Mignemi, \PL{B672}, 186 (2009).
\ref [9] E.J. Hellund and K. Tanaka, \PR{94}, 192 (1954);
 G. Jaroszkiewicz, \JoP{A28}, L343 (1995);
 J.M. Romero and A. Zamora, \PR{D70}, 105006 (2004);
 E.R. Livine and D. Oriti, \JHEP{0406}, 050 (2004);
 R. Banerjee, S. Kulkarni and S. Samanta, \JHEP{0605}, 077 (2006);
 M.V. Battisti and S. Meljanac, \PR{D79}, 067505 (2009).
\ref [10] J. Kowalski-Glikman and L. Smolin, \PR{D70}, 065020 (2004).
\ref [11] C. Chryssomakolos and E. Okon, \IJMP{D13}, 1817 (2004).
\ref [12] C.N. Yang, \PR{72}, 874 (1947).
\ref [13] S. Mignemi, \CQG{26}, 245020 (2009).
\ref [14] S. Mignemi, Annalen der Physik {\bf522}, 924 (2010).
\ref [15] H.G. Guo, C.G. Huang and H.T. Wu, \PL{B663}, 270 (2008);
 M.C. Carrisi and S. Mignemi, \PR{D82}, 105031 (2010);
 R. Banerjee, K. Kumar and D. Roychowdhury, \JHEP{1103}, 060 (2011).
\ref [16] S. Mignemi, \PR{D84}, 025021 (2011).
\ref [17] Lei Lu and A. Stern, \arx{1108.1832}.
\ref [18] A. Kempf, G. Mangano and R.B. Mann, \PR{D52}, 1108 (1995).
\ref [19] J. Magueijo and L. Smolin, \CQG{21}, 1725 (2004).
\ref [20] M. Abramowitz and I.A. Stegun, {\sl Handbook of mathematical fuctions},
Dover 1965.

\endref
\end